\documentclass[prl, twocolumn, preprintnumbers, superscriptaddress, nofootinbib]{revtex4}
\pdfoutput=1 
\usepackage{amssymb,amsmath}
\usepackage{color}
\usepackage{graphicx}
\usepackage{multirow}

\begin{document}

\preprint{JLAB-THY-10-1291 / TCDMATH-10-09}

\title{The phase-shift of isospin-2 $\pi\pi$ scattering from lattice QCD}

\author{Jozef J. Dudek}
\email{dudek@jlab.org}
\affiliation{Jefferson Laboratory, 12000 Jefferson Avenue,  Newport News, VA 23606, USA}
\affiliation{Department of Physics, Old Dominion University, Norfolk, VA 23529, USA}

\author{Robert G. Edwards}
\affiliation{Jefferson Laboratory, 12000 Jefferson Avenue,  Newport News, VA 23606, USA}

\author{Michael J. Peardon}
\affiliation{School of Mathematics, Trinity College, Dublin 2, Ireland}

\author{David G. Richards}
\affiliation{Jefferson Laboratory, 12000 Jefferson Avenue,  Newport News, VA 23606, USA}

\author{Christopher E. Thomas}
\affiliation{Jefferson Laboratory, 12000 Jefferson Avenue,  Newport News, VA 23606, USA}

\collaboration{for the Hadron Spectrum Collaboration}

\begin{abstract}
Finite-volume lattice QCD calculations offer the possibility of extracting resonance parameters from the energy-dependent elastic phase-shift computed using the L\"uscher technique. In this letter, as a trial of the method, we report on the extraction of the non-resonant phase-shift for $S$ and $D$-wave $\pi\pi$ isospin-2 scattering from dynamical lattice QCD computations. We define a variational basis of operators resembling pairs of pions of definite relative momentum and extract a spectrum of excited states that maps to phase-shifts at a set of discrete scattering momenta. Computations are performed with pion masses between $400$ and $520$ MeV on multiple spatial volumes. We observe no significant quark mass dependence in the phase-shifts extracted which are in reasonable agreement with the available experimental data at low momentum. 
\end{abstract}


\maketitle 


\paragraph{Introduction:}
The hadron spectrum and interactions of QCD can be studied from first
principles using numerical simulation of the quark and gluon
fields on a finite lattice.
While significant progress has been made in
studying isolated excited meson states with $q\bar{q}$-like operators
\cite{Dudek:2009qf, Dudek:2010wm}, it remains challenging to extract
properties of resonances that appear in the scattering of stable
hadrons. One procedure, due to L\"uscher \cite{Luscher:1990ux}, maps
the discrete spectrum of eigenstates of QCD in a finite cubic volume
to the phase shift for elastic scattering. By extracting multiple
excited eigenstates within a given quantum number sector, one can map
out the phase shift as a function of scattering momentum and, if present
in that channel, observe resonant behaviour.

In this letter, we demonstrate the feasibility of the technique in a
simple sector, that of $\pi\pi$ scattering in isospin-2 ($I=2$), where the
interaction is not strong enough to form a resonance, but rather is
weak and repulsive. For the first time using this method, we extract the
$S$- and $D$-wave phase shifts as a function of scattering
momentum. This procedure is carried out independently on multiple
volumes to validate the finite-volume method.
We find through computations at a range of quark masses that at the level of precision
attained the phase shift is largely quark mass independent.

Previous lattice QCD calculations of $\pi\pi$ scattering have limited themselves to extracting the phase shift at near-zero energy, 
more conveniently expressed via the scattering length 
\cite{Beane:2007xs, Feng:2009ij}, or by performing the same calculation in a moving frame, a single phase shift point at non-zero energy \cite{Sasaki:2008sv}. 

In contrast, we use the
``distillation" method \cite{Peardon:2009gh} to construct both creation
and annihilation operators of definite 
$\pi\pi$ relative momentum, and employ
them to form a variational basis of composite QCD
operators that resemble pairs of pions.
This enables us to extract a spectrum of multiple states with $I=2$, $\ell^P=0^+,2^+$ 
($\ell$ is the partial wave and $P$ the parity) 
and, using the L\"uscher technique, we find the phase shift
as a discrete function of the scattering momentum. This sets the groundwork
for investigating resonances in other meson-meson scattering channels.

Experimentally, $\pi\pi$ $I=2$ phase shifts have been extracted from $\pi N \to \pi \pi N'$ charge-exchange scattering reactions, treating the dominantly-exchanged pion as approximately on-shell owing to the proximity of the $t$-channel pole to the physical small-$t$ region. The extant data \cite{Hoogland:1977kt, Cohen:1973yx, Losty:1973et, Durusoy:1973aj}  for $\delta_{\ell=0}$ and  $\delta_{\ell=2}$ are broadly consistent 
in the low-energy region measured and there is little statistically significant evidence for inelasticity.\\

\paragraph{Finite volume analysis:}

L\"uscher's method relates the discrete spectrum of energy levels in a finite volume to phase shifts evaluated at the scattering momenta corresponding to the extracted energy values. Complications arise from the cubic symmetry of the lattice boundary which reduces the irreducible symmetry channels from the set of all integer spins to a finite set of irreducible representations. The relevant \emph{irreps}, $\Gamma$, for $\pi\pi$ isospin-2 scattering at low momentum are $A_1^+$ which contains continuum spins $\ell=0,4\ldots$, $T_2^+(\ell=2,4\ldots)$, $E^+(\ell=2,4\ldots)$ and $T_1^+(\ell=4\ldots)$. Odd $\ell$ do not contribute due to Bose symmetry.

Once the finite volume energy levels, $E_{\pi\pi}$ are obtained from an explicit Monte-Carlo calculation on a fixed volume ($L^3$) lattice, the scattering momenta follow assuming a continuum-like dispersion relation, $k = \sqrt{ (E_{\pi\pi}/2)^2 - m_\pi^2 }$. The desired phase-shifts are embedded in an equation
\begin{equation}
	\det\left[  e^{2i \boldsymbol{\delta}(k)} - \mathbf{U}_\Gamma\left(k\tfrac{L}{2\pi}\right)
 \right] = 0, \label{lu} 
\end{equation}
where $\mathbf{U}_\Gamma\left(k\tfrac{L}{2\pi}\right)$ is a matrix in the space of partial waves, $\ell$, of known functions particular to this irrep, $\Gamma$, evaluated at the
scattering momentum, $k$. $e^{2i \boldsymbol{\delta}(k)}$ is a
diagonal matrix featuring phase-shifts, $\delta_\ell(k)$, for all partial waves
contributing to the irrep $\Gamma$. The dimension of these matrices is
formally infinite, since there are an infinite number of possible
partial waves contributing to each irrep $\Gamma$. However, one can argue
that, since higher waves typically contribute less at low
momentum, one can cut-off the dimension at some
low-spin and, provided the results are reasonably insensitive to this
cutoff, reliably extract the phase shift for low partial waves. This is
the core of the L\"uscher method \cite{Luscher:1990ux}.

The aim then is to solve for some set of phase-shifts $\{\delta_\ell(k)\}$, but since Equation \ref{lu} is one equation (per energy level) in several unknowns, this will not be possible. Instead we will attempt to bound the size of all $\delta_\ell(k)$ for $\ell$ higher than the lowest in each irrep using other information. In practice we will assume that, in the energy region accessible to us, $\delta_{\ell>4} \approx 0$ and consider only the effect of a non-zero $\delta_4(k)$.\\

\paragraph{Correlator construction and variational analysis:}

In order to obtain the finite-volume energy spectrum, we form a matrix
of correlators using a basis of $\pi\pi$ operators that is then
diagonalised using the variational method \cite{Luscher:1990ck,
  Michael:1985ne, Dudek:2010wm}. The operators are constructed to
resemble a pair of pions with total momentum zero and definite
relative momentum:
\begin{equation}
	{\cal O}_{\pi\pi}^{\Gamma, \gamma}(|\vec{p}|) = \sum_{m} {\cal S}^{\ell, m}_{\Gamma, \gamma} \sum_{\hat{p}} Y_\ell^m(\hat{p})\,{\cal O}_\pi(\vec{p}) {\cal O}_\pi(-\vec{p}). \nonumber
\end{equation}
The \emph{subduction} coefficients, ${\cal S}^{\ell, m}_{\Gamma,
  \gamma}$, project operators of definite $\ell$ into definite
\emph{irreps}, $\Gamma$ - their explicit forms can be found in
Appendix A of \cite{Dudek:2010wm}. The sum over directions of momentum, $\hat{p}$,
at a fixed magnitude is limited to those allowed by
the periodic cubic boundary conditions.  On a lattice with spatial extent $L$ these are
$\vec{p} = \tfrac{2\pi}{L}\vec{n}$ for a vector of integers $\vec{n}$.

In this first study we utilise only a simple operator capable of interpolating a pion at momentum $\vec{p}$ from the vacuum, 
\begin{equation}
{\cal O}_\pi(\vec{p}) = \sum_{\vec{x}} e^{i \vec{p}\cdot \vec{x}} \left[\bar{\psi} \Box_\sigma \gamma^5 \Box_\sigma \psi\right](\vec{x}), \nonumber
\end{equation}
where the quark fields are acted upon by a distillation smearing
operator that emphasises the low momentum quark and gluon modes that
dominate low mass hadrons. In this study we use $\Box_\sigma =
\sum_{n=1}^{N_\mathrm{vecs}} e^{\sigma^2 \lambda_n/4} \xi_n
\xi^\dag_n$ where $\lambda_n, \xi_n$ are the eigenvalues and
eigenvectors of the gauge-covariant three-dimensional Laplacian
operator (see \cite{Peardon:2009gh, Dudek:2010wm} for details; 
$\sigma=0$ was used in \cite{Dudek:2009qf,Dudek:2010wm}).
It is distillation that factorises the construction of correlators in such a
way as to make possible the projection onto definite inter-pion
momentum at both source and sink, something that is not possible in
the traditional ``point-all" method. Details of the distillation correlator construction can be found in \cite{Peardon:2009gh}.

Our variational basis in the irrep $A_1^+$ consists of operators with
$|\vec{p}|^2 = \left(\tfrac{2\pi}{L}\right)^2 (0,1, \ldots 4)$ each with
two smearing radii $\sigma = 0.0, 4.0$, giving a ten dimensional
basis. For $E^+$ we have $|\vec{p}|^2 = \left(\tfrac{2\pi}{L}\right)^2
(1,2,4)$ and two smearings and $T_2^+$ with $|\vec{p}|^2 =
\left(\tfrac{2\pi}{L}\right)^2 (2,3)$ and two smearings. The $T_1^+$ irrep has lowest spin $\ell=4$ for two pions. However the lowest momentum from which a $T_1^+$ operator can be constructed is $|\vec{p}|^2 = 5 \left(\tfrac{2\pi}{L}\right)^2 $ and this is the only one we used. With these operators at source and sink, we form all correlators using Wick
contractions relevant for $I=2$.

Computations are performed on anisotropic lattices with three dynamical flavors of Clover fermions with spatial lattice spacing $a_s \sim 0.12 \,\mathrm{fm}$ and finer temporal spacing, $a_t^{-1} \sim 5.6 \, \mathrm{GeV}$, see Table \ref{tab:lattices} and \cite{Lin:2008pr}. A precise measure of the anisotropy, $\xi = a_s/a_t$, is required to determine the spatial length of the lattice in temporal lattice units, $L/a_t = \xi L/a_s$. Fitting single-pion correlators at finite momentum, $a_s\vec{p} = \tfrac{2\pi}{L/a_s}\vec{n}$, determines $a_t E_\pi(|\vec{n}|)$ and $\xi$ follows from fitting the dispersion relation 
\begin{equation}
\big(a_t E_\pi(|\vec{n}|)\big)^2 = (a_t m_\pi)^2 + \tfrac{1}{\xi^2} \left(\tfrac{2\pi}{L/a_s} \right)^2 |\vec{n}|^2, \nonumber
\end{equation} 
for multiple values of $|\vec{n}|$ and $L/a_s$. Explicitly we find $\xi = 3.459(4), 3.454(5), 3.459(3)$ on respectively the $m_\pi = 396, 444, 524$ MeV lattices, showing the lack of quark mass dependence observed previously and utilised in the dynamical tuning of the lattice action\cite{Edwards:2008ja}. Mass-dimension quantities multiplied by the temporal lattice spacing, $a_t$, are scale-set using the procedure outlined in \cite{Dudek:2010wm}, using the $\Omega$-baryon mass determined on the same lattice, $m = \frac{a_t m}{a_t m_\Omega}\cdot m_\Omega^{\mathrm{phys.}}$. The continuum scaling of the results is not investigated in this calculation at a single lattice spacing.

\begin{table}
\begin{tabular}{c|cccc}
$m_\pi/\mathrm{MeV}$ &  $(L/a_s)^3 \times (T/a_t)$   &$N_{\mathrm{cfgs}}$ & $N_{\mathrm{t_{srcs}}}$ & $N_{\mathrm{vecs}}$ \\
\hline\hline
\multirow{2}{*}{$524$} & $16^3 \times 128$ & $496$ & $4$ &$64$\\
                       & $20^3 \times 128$ & $377$ & $4$ &$96$\\
                       \hline
\multirow{2}{*}{$444$} & $16^3 \times 128$ & $605$ & $5$ &$64$\\
                       & $20^3 \times 128$ & $321$ & $3$ &$128$\\
                       \hline
\multirow{3}{*}{$396$} & $16^3 \times 128$ & $439$ & $16$ & $64$\\
                       & $20^3 \times 128$ & $535$ & $3$ & $128$\\
                       & $24^3 \times 128$ & $548$ & $4$ & $162$\\	
	\end{tabular}
\caption{Lattices used in this study. $N_\mathrm{vecs}$ indicates the number of eigenvectors of the laplacian used in the distillation method.}
\label{tab:lattices}
\end{table}

\begin{figure}
 \centering
\includegraphics[width=0.49\textwidth,bb=40 0 650 410]{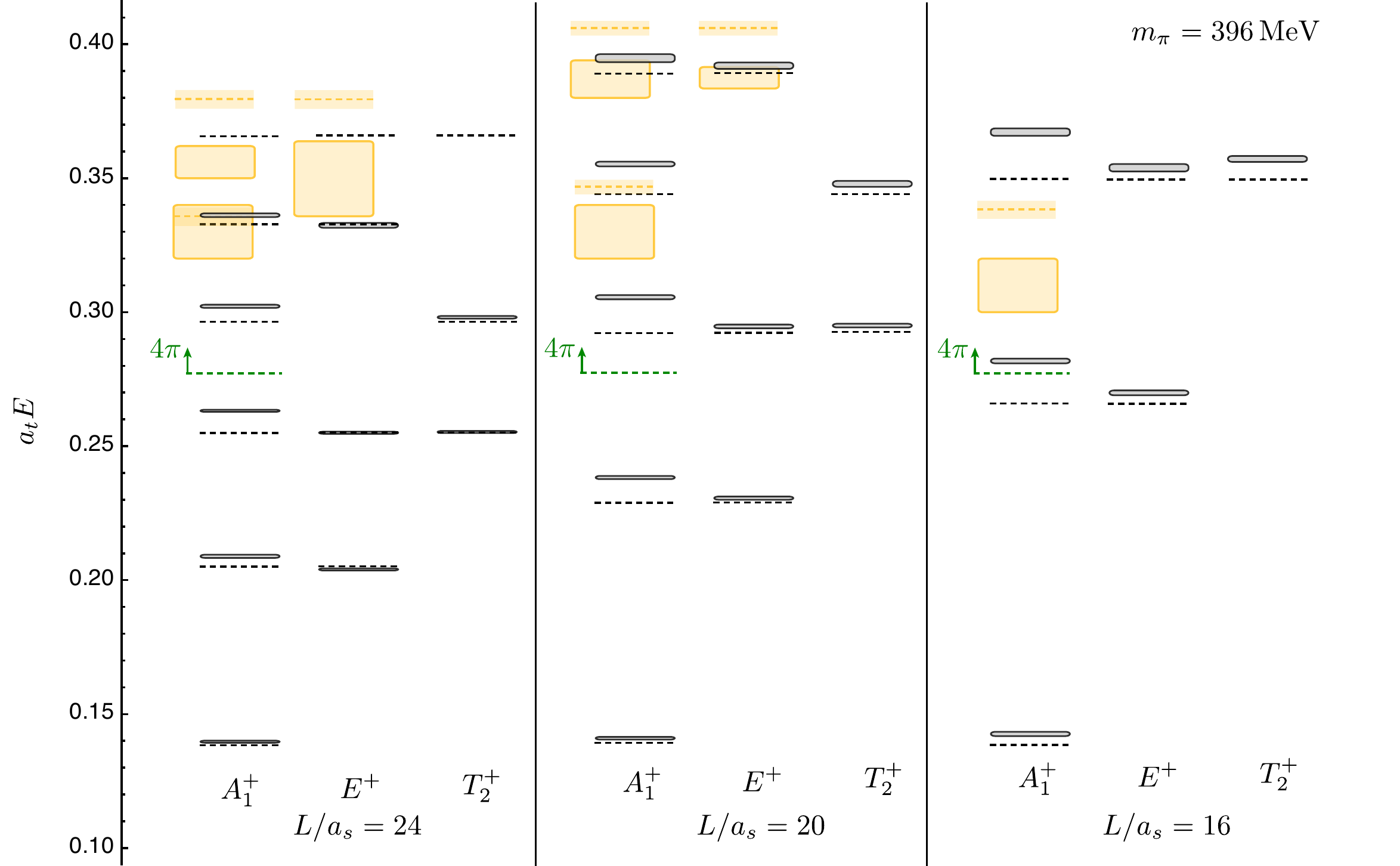} 
\caption{Low-lying spectrum, in units of the temporal lattice spacing, of finite volume states in irreps $A_1^+, E^+, T_2^+$ at $m_\pi = 396\,\mathrm{MeV}$ for $L/a_s=16,20,24$. The box height indicates the statistical uncertainty on the energy. Orange boxes correspond to states suspected of being $\pi \pi^\star$ scattering states. Dashed lines indicate the non-interacting energy of pion pairs with the allowed lattice momenta between them.  \label{spectrum}}
\end{figure}

In Figure \ref{spectrum} we show the finite-volume spectra obtained
with $m_\pi = 396\,\mathrm{MeV}$ and $L/a_s=16, 20, 24$. We clearly
observe shifts relative to the energy of two non-interacting pions
with back-to-back momentum of $\vec{p} =\tfrac{2\pi}{L}\vec{n}$, $E_{\pi\pi} = 2 \sqrt{m_\pi^2 + |\vec{p}|^2}$. It is this energy shift that L\"uscher's method
relates to the scattering phase shift through Equation \ref{lu}. It is the fact that we are able to resolve excited energy levels with a statistical precision below 1\% that makes possible an extraction of the scattering phase shift as a function of scattering momentum.

\begin{figure}
 \centering
\includegraphics[width=0.5\textwidth,bb=20 20 380 260]{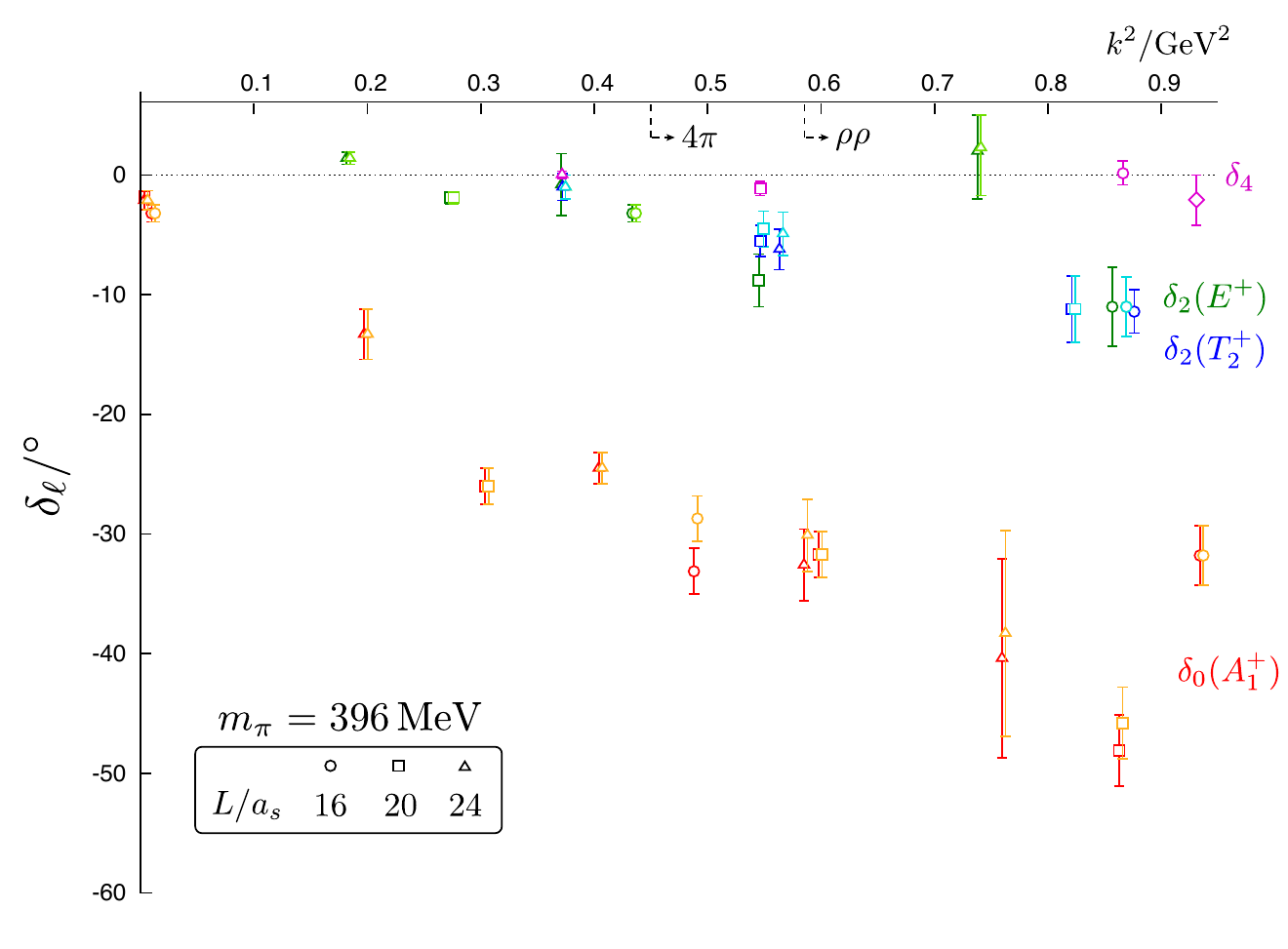} 
\caption{Phase-shifts extracted from spectra with $m_\pi = 396\,\mathrm{MeV}$. Red ($A_1^+$), green ($E^+$), blue ($T_2^+$) colored points assume $\delta_4=0$; orange ($A_1^+$), light green ($E^+$), cyan ($T_2^+$) colored points (shifted slightly to the right) used estimated $\delta_4$ as described in the text: note that the corrected $\delta_2$ values from $E^+,T_2^+$ coincide by construction at momenta near $|\vec{p}|^2 = 2 \cdot \left(\tfrac{2\pi}{L}\right)^2$ . Estimated $\delta_4$ shown by pink points. Also indicated are the positions of inelastic thresholds into $4\pi$ and $\rho\rho$.   \label{840}}
\end{figure}

\paragraph{Phase shift:}
For each $\pi\pi$ energy level in each irrep in Figure \ref{spectrum}
we can set up an Equation \ref{lu} to be solved for the phase shifts,
$\delta_\ell$. The simplest way to solve these equations is to neglect
the contribution of $\ell \geq 4$ to obtain $\delta_0$ from
$A_1^+$ and $\delta_2$ from $T_2^+$ or $E^+$. Doing so gives the red, green and blue colored points in Figure \ref{840}. The small discrepancies between $E^+$ and $T_2^+$ extractions of $\delta_2$ at $k^2 \sim 0.35, 0.55, 0.85\, \mathrm{GeV}^2$ (corresponding to the levels at $a_t E \sim 0.25, 0.29, 0.36$ in Figure \ref{spectrum}), have a possible origin in the neglect of a non-negligible value of $\delta_4$. 
We can estimate the size of this $\delta_4$ by solving the coupled system of Equations \ref{lu} for $T_2^+$ and $E^+$ at the relevant energy for the two unknowns, $\delta_2, \delta_4$. The values of $\delta_4$ so extracted are shown by the pink points in Figure \ref{840}. 

For a direct estimate of $\delta_4$ from $T_1^+$, only the $L/a_s=24$ lattice has a point within our plotted range of scattering momentum. The extracted point is shown by the pink diamond in Figure 2, and is in good agreement with the other estimates, showing that $|\delta_4|$ is less than $2^\circ$ over the whole of the explored momentum range.

\begin{figure}
 \centering
\includegraphics[width=0.5\textwidth,bb=20 0 380 200]{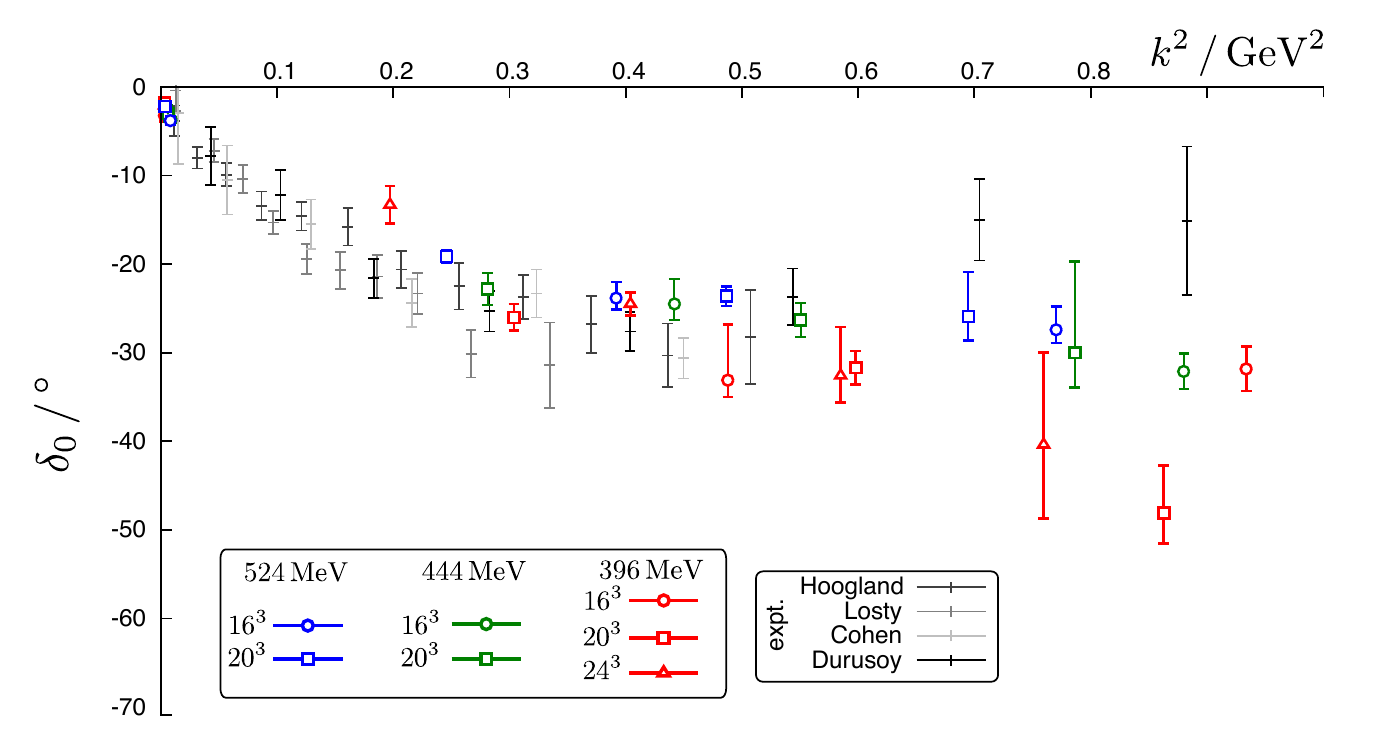} 
\includegraphics[width=0.5\textwidth,bb=20 20 380 250]{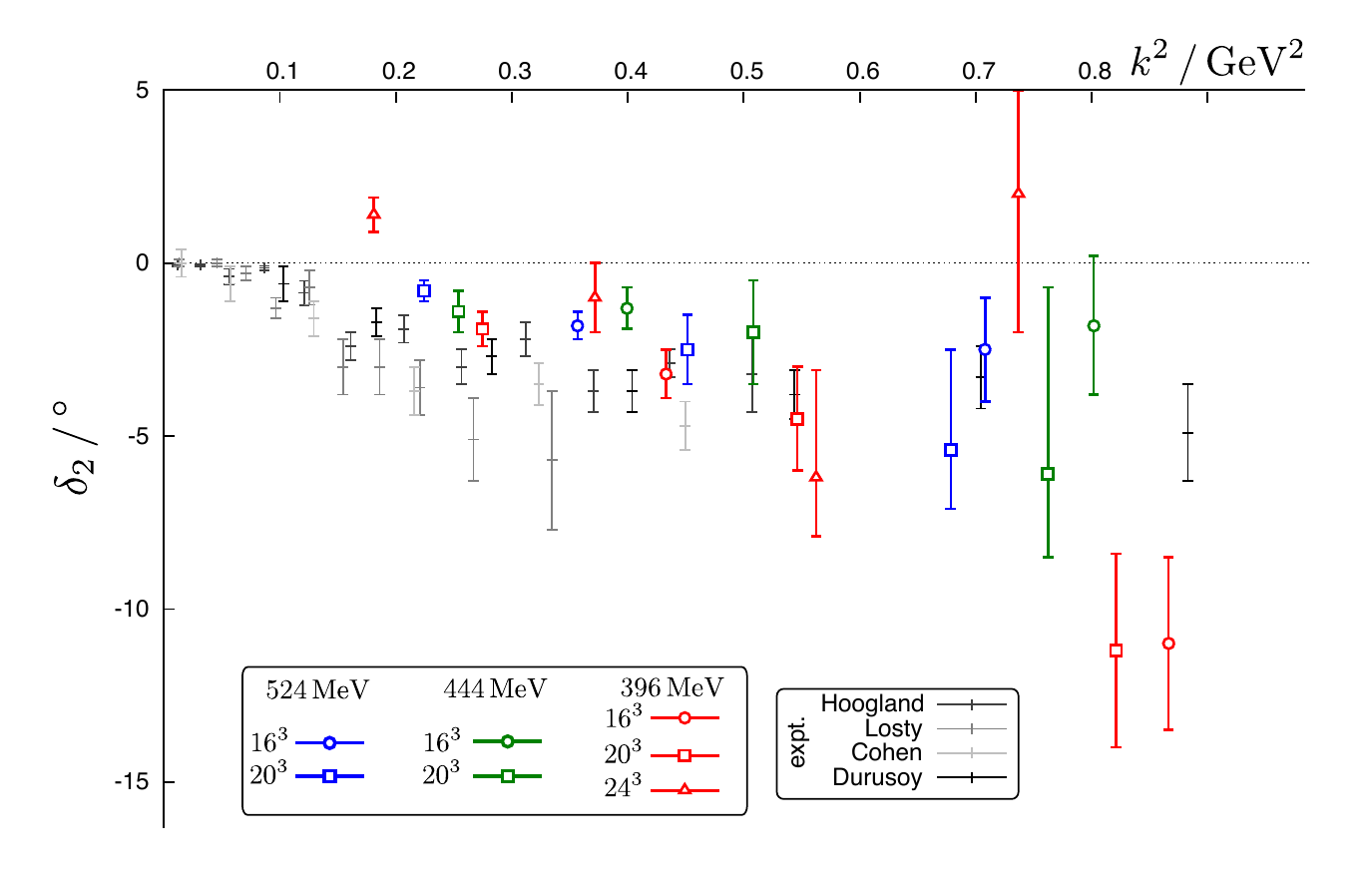} 
\caption{Phase-shift in degrees for $\pi\pi$ $I=2$ scattering with $\ell=0$($\delta_0$) and $\ell=2$($\delta_2$). Lattice results at various pion masses and volumes. Experimental data from \cite{Hoogland:1977kt, Cohen:1973yx, Losty:1973et, Durusoy:1973aj}.  \label{delta02}}
\end{figure}

With an estimated magnitude of $\delta_4(k)$ in hand (from interpolation between the determined points), we can solve Equation \ref{lu} including the effect of the $\ell = 4$ wave. This gives rise to the orange, light green and cyan colored points in Figure \ref{840} which are seen to differ relatively little from the points with $\delta_4$ assumed to be zero. For final presentation we enlarge the errorbar to include the effect of the estimated $\delta_4$ giving rise to asymmetric errorbars in Figure \ref{delta02}.

As indicated in Figure \ref{840}, the $4\pi$ threshold opens within the energy range of our extracted phase-shifts and technically for energies above this the 
formalism leading to Equation \ref{lu} is not rigorously correct. On the other hand, there is relatively little evidence experimentally for considerable inelasticity in the $\pi \pi$ isospin-2 channel in the energy range so-far probed - what little data there is does not show statistically significant deviation from an elastic approximation \cite{Cohen:1973yx, Losty:1973et}. As an initial approximation, we shall assume that the inelasticity is negligible and continue to use Equation \ref{lu} above the inelastic threshold. The $\ell=2$ phase-shift extracted from $E^+$, $T_2^+$ irreps should be less sensitive to any inelasticity since the effective threshold in finite-volume is higher as it requires at least one unit of relative momentum in the $4\pi$ system. 
Future calculations should test the elasticity assumption by computing correlators using operators that resemble four pions projected into isospin-2 in the appropriate partial waves. \\

\paragraph{Results:}

In Figure \ref{delta02} we show our results for $S$ and $D$-wave phase shifts, at a range of pion masses, along with experimental data taken from  \cite{Hoogland:1977kt, Cohen:1973yx, Losty:1973et, Durusoy:1973aj}. We observe reasonable agreement with the experimental data at lower scattering momenta, where the scattering is purely elastic, for all the pion masses computed. 
This suggests that it is possible that the phase-shift is only mildly dependent upon pion mass. Of course, one requires lattice computations at smaller pion masses to verify that the agreement with experiment continues.

\begin{figure}[!b]
 \centering
\includegraphics[width=0.49\textwidth,bb=30 15 360 180]{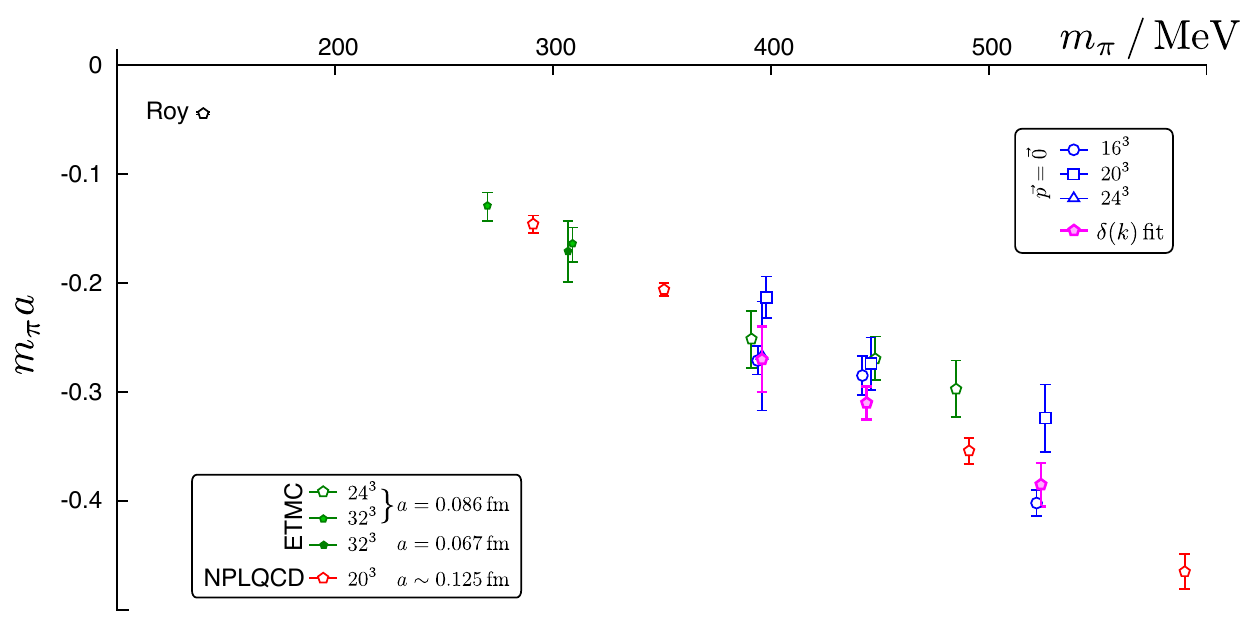} 
\caption{$S$-wave scattering length. Blue points from $\vec{p}=\vec{0}$ correlators, pink points from effective range fits to $\delta(k)$. Comparison to lattice results of \cite{Beane:2007xs, Feng:2009ij} and Roy equations analysis of experimental data \cite{Colangelo:2001df}. \label{scatlen}}
\end{figure}

Using only $\pi\pi$ and $\pi$ correlators at zero momentum we can perform the extraction of the scattering length using the methodology of \cite{Beane:2007xs, Feng:2009ij}. The scattering lengths so obtained are shown in Figure \ref{scatlen}, where they are seen to be in reasonable agreement with the precision data of \cite{Beane:2007xs}, computed on a lattice of similar spatial lattice spacing. 

We can also obtain estimates for the scattering length and effective range by fitting the $k$ dependence of $\delta_0(k)$, where we find that scattering lengths largely agree with the estimates from the simple method above, while the effective range is small but only poorly determined.\\

\paragraph{Summary and prospects:}
We have demonstrated the feasibility of an explicit application of the L\"uscher finite-volume framework in dynamical lattice QCD. Using multiple excited state energy levels extracted in a single volume, we have determined the $S$- and $D$-wave $\pi\pi$ isospin-2 phase shifts as a function of scattering momentum. Multiple volumes are then used for validation and estimation of the effect of neglected higher partial waves. We estimate that $|\delta_4 |\lesssim 2^\circ$ for $k < 1\,\mathrm{GeV}$.

We observe no significant pion mass dependence in the phase shift below $k \sim 1\, \mathrm{GeV}$, with results for $m_\pi \gtrsim 400 \, \mathrm{MeV}$ being in reasonable agreement with experimental data at low scattering momentum. For \emph{precision} coverage of experimentally relevant kinematics, we would require still larger volumes to sample points at smaller scattering momentum in the elastic region. 

This calculation sets the groundwork for an investigation of the
resonances in meson-meson scattering that arise from the strong
interaction.  Inclusion of quark annihilation diagrams in the calculation
of correlators \cite{Dudek:2011tt} will enable the $I=1$ $\pi\pi$ sector to be studied,
where one expects to see the $\rho$ resonance appearing as a rapidly rising phase shift. Some attempts in
this direction have been made~\cite{Gockeler:2008kc, Aoki:2007rd, Feng:2010es}, but
using only a small basis of operators and subsequently extracting a
very limited number of phase-shift points. \emph{Distillation}
and stochastic variants~\cite{Peardon:2009gh, Morningstar:2010ae, Bulava:2010em} will
allow us to efficiently construct a large basis and thus map out many
points on the phase-shift curve. In future work, we will explore a range
of different scattering hadrons in various partial-waves.\\

\paragraph{Acknowledgments - }
We thank our colleagues within the Hadron Spectrum Collaboration. Chroma~\cite{Edwards:2004sx} and GPU-code from Clark et al.~\cite{Clark:2009wm, Babich:2010mu} were used to perform this work 
on clusters at Jefferson Laboratory using time awarded under the USQCD 
Initiative. Authored by Jefferson Science Associates, LLC under U.S. DOE Contract No.
DE-AC05-06OR23177. MP is supported by Science Foundation Ireland under research grant 07/RFP/PHYF168


\bibliography{pipi}

\end{document}